\documentclass[runningheads]{llncs}
\usepackage{graphicx}
\usepackage{mathptmx} 
\usepackage{times} 
\usepackage{amssymb}  
\usepackage{amsmath} 
\usepackage{booktabs}
\usepackage{multirow} 
\usepackage{multicol}
\usepackage{enumitem}
\usepackage{tabularx}
\usepackage{caption}
\usepackage{marvosym}
\usepackage[colorlinks=true, linkcolor=blue, citecolor=blue, urlcolor=blue]{hyperref}
\begin{document}
\title{
Robust Real-Time Endoscopic Stereo Matching under Fuzzy Tissue Boundaries }
%
%
\author{Yang Ding\inst{1} \and Can Han\inst{1} \and Sijia Du\inst{1} \and Yaqi Wang\inst{2} \and Dahong Qian\inst{1}\textsuperscript{(\Letter)}}
%
\institute{The School of Biomedical Engineering, Shanghai Jiao Tong University, Shanghai 200240, China\\ \email{dahong.qian@sjtu.edu.cn}   \and
Innovation Center for Electronic Design Automation Technology, Hangzhou Dianzi University, Hangzhou 310018, China
}
%
\maketitle              
\begin{abstract}
Real-time acquisition of accurate scene depth is essential for automated robotic minimally invasive surgery. Stereo matching with binocular endoscopy can provide this depth information. However, existing stereo matching methods, designed primarily for natural images, often struggle with endoscopic images due to fuzzy tissue boundaries and typically fail to meet real-time requirements for high-resolution endoscopic image inputs. To address these challenges, we propose \textbf{RRESM}, a real-time stereo matching method tailored for endoscopic images. Our approach integrates a 3D Mamba Coordinate Attention module that enhances cost aggregation through position-sensitive attention maps and long-range spatial dependency modeling via the Mamba block, generating a robust cost volume without substantial computational overhead. Additionally, we introduce a High-Frequency Disparity Optimization module that refines disparity predictions near tissue boundaries by amplifying high-frequency details in the wavelet domain. Evaluations on the SCARED and SERV-CT datasets demonstrate state-of-the-art matching accuracy with a real-time inference speed of 42 FPS. The code is available at https://github.com/Sonne-Ding/RRESM.

\end{abstract}

\section{Introduction}

Minimally invasive surgery (MIS) has become a preferred surgical approach due to its reduced invasiveness and faster recovery times~\cite{c38}. Endoscopy can provide essential visual guidance in MIS. Although endoscopy provides essential visual guidance in MIS, it faces inherent limitations—such as a restricted field of view, lack of tactile feedback, and diminished spatial awareness~\cite{c3,c35}. To overcome these challenges, computer-assisted intervention techniques have been developed to extract spatial depth information from endoscopic images, with depth estimation emerging as a key focus~\cite{c35}. 

Endoscopic stereo matching is the method to obtain tissue depth information from binocular images~\cite{c1,c2}. Specifically, it generates a disparity map that can be converted into a depth map through a simple mapping relationship. Although there have been numerous studies on stereo matching in the domain of natural images, endoscopic stereo matching still faces some challenges due to its unique application scenarios. Tissue boundaries in endoscopic images are blurred by low-contrast, gradual transitions~\cite{c21}, making accurate depth estimation difficult. Meanwhile, the increasing adoption of high-definition endoscopes places additional burden on stereo matching to preserve computational efficiency and real-time performance. Therefore, developing an efficient, accurate, and robust endoscopic stereo matching method is essential.

\begin{figure}[h]
  \centering
  \includegraphics[width=0.9\textwidth]{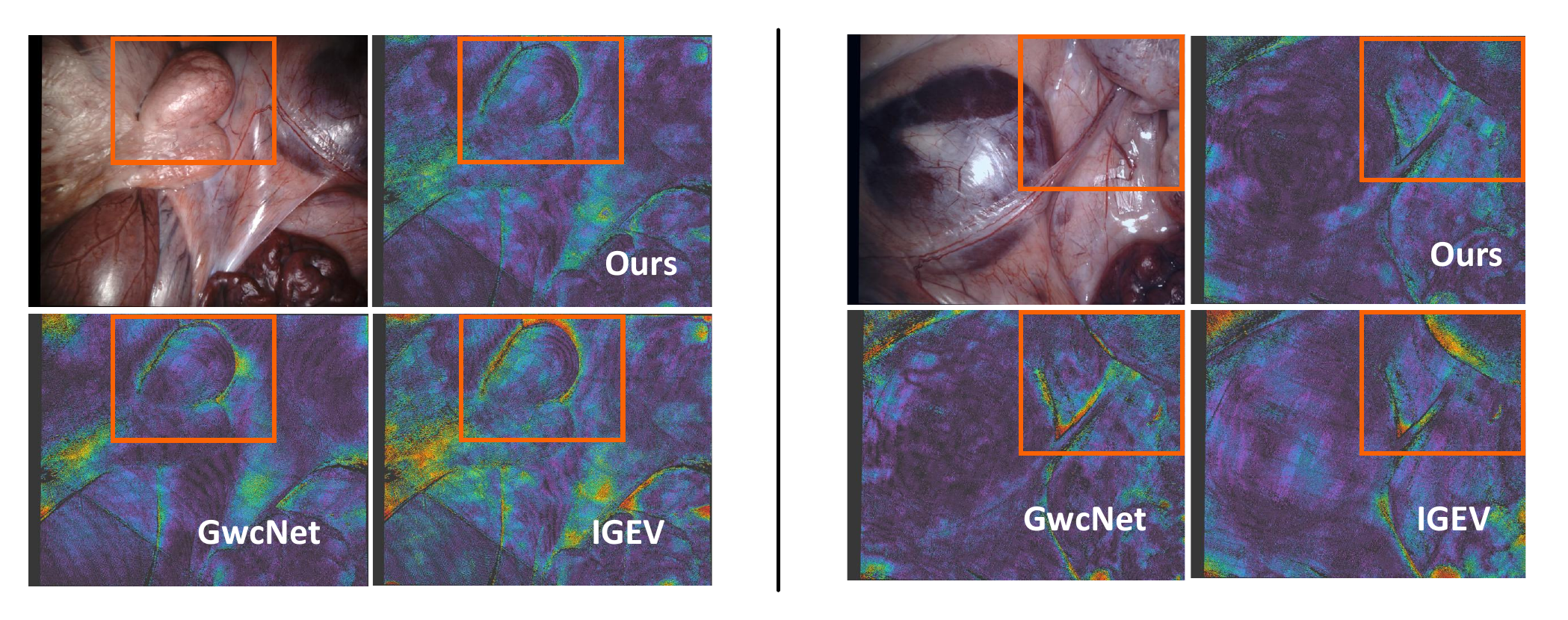}
    \caption{Depth error maps near fuzzy boundaries on the SCARED dataset. Brighter regions indicate larger depth errors. Our method performs well in estimating depth around fuzzy tissue boundaries, outperforming the state-of-the-art natural image methods (e.g., GwcNet and IGEV).}
    \vspace{-0.5cm}
  \label{fig:intro}
\end{figure}

Since Zbontar et al.~\cite{c4} first introduced convolutional neural networks (CNNs) into stereo matching, numerous deep learning-based methods have been proposed in this field. Compared to traditional approaches, deep models—leveraging complex architectures, adaptive feature extraction, and learned optimization—have achieved significantly higher matching accuracy~\cite{c12,c14,c15,c16}. 
Currently, stereo matching methods primarily fall into two categories: those based on 3D convolutions~\cite{c1,c4,c7,c8,c17,c26}, and those based on iterative optimization~\cite{c10,c18,c19,c28}. 3D convolution-based approaches effectively encode geometric information but incur substantial computational overhead~\cite{c5}. Iterative optimization methods, typically employing recurrent neural networks (RNNs) to refine disparity estimations over multiple steps~\cite{c18}, offer a trade-off between accuracy and efficiency by adjusting the number of iterations. However, they still exhibit high latency when processing high-resolution endoscopic images.
Moreover, most of these models are designed for natural image domains and generalize poorly to endoscopic scenes. As shown in Fig.~\ref{fig:intro}, both the 3D CNN-based GwcNet~\cite{c8} and the iterative method IGEV~\cite{c18} produce large depth errors near fuzzy tissue boundaries.

To enable real-time stereo matching for robotic applications in endoscopic scenes with fuzzy tissue boundaries, we propose RRESM, a lightweight and efficient stereo matching framework. RRESM employs MobileNetV4~\cite{c22} as a compact feature extractor to reduce computational overhead while preserving representational capacity.
For cost aggregation, we introduce a novel attention-based module, the \textbf{3D Mamba Coordinate Attention (MCA) Module}. Instead of computing attention over the entire 3D cost volume, MCA performs axis-wise attention in 1D, significantly reducing computational complexity. It further leverages the Mamba block~\cite{dao2024transformers} to model long-range dependencies along all three spatial dimensions. This design enables effective cost aggregation with minimal computational overhead. Experimental results show that MCA significantly outperforms traditional architectures such as the Stacked Hourglass Network~\cite{c21}, which relies heavily on stacked 3D convolutions.
To further refine disparity maps, especially around tissue boundaries, we propose the \textbf{High-Frequency Disparity Optimization (HFDO)} module. This module uses the Haar wavelet transform to decompose contextual features into low- and high-frequency components. It enhances high-frequency signals and reconstructs features via inverse wavelet transform, thereby enriching the disparity representation with fine structural details. The refined features are then projected into disparity space to enhance depth estimation in high-frequency regions. Our method demonstrates superior performance in depth estimation near fuzzy tissue boundaries, as shown in Fig.\ref{fig:intro}.
We evaluate RRESM on the SCARED~\cite{c20} and SERV-CT~\cite{c29} datasets, demonstrating both accuracy and real-time performance. Our method achieves the best average MAE of \(2.592\,\text{mm}\) on the SCARED dataset, with an inference time of \(23.38\,\text{ms}\) per frame at a resolution of \(1024 \times 1280\).


\section{Related Works}
\subsection{Fuzzy Boundaries Optimization in Stereo Matching}
Boundary regions in images are typically characterized by high-frequency components, which pose challenges for accurate stereo matching. Several recent works have focused on improving disparity estimation in such regions by enhancing high-frequency detail prediction~\cite{c10,c31,c32}. DLNR~\cite{c10} points out that the tight coupling between the update matrix and hidden state transition in the GRU module used by RAFT can lead to degraded performance in high-frequency regions. To mitigate this, DLNR replaces GRUs with LSTMs (Long Short-Term Memory), achieving improved accuracy in disparity prediction near edges.Selective-Stereo~\cite{c31} further observes that GRUs with fixed receptive fields are limited in their ability to capture both high-frequency edge information and low-frequency texture information. To address this, it introduces the Selective Recurrent Unit (SRU), which incorporates multi-scale receptive fields into the GRU framework. This design allows the network to adaptively process features across multiple frequency bands, enhancing disparity estimation performance in both boundary and low-frequency texture regions.

\subsection{Real-time Stereo Matching}
For real-time stereo matching, there are primarily two categories: CNN-based cost aggregation methods and iterative optimization-based methods The use of CNNs in stereo matching traces back to the nascent stages of deep learning. These methods predominantly utilize CNNs to extract features and construct a cost volume for disparity estimation. CNN-based methods can be divided into 2D and 3D architectures contingent upon the processing of the cost volume. For instance, 2D architectures like DispNet~\cite{c23} employ 2D convolutions to process the cost volume, thereby achieving real-time performance albeit with relatively lower accuracy. Conversely, 3D architectures, such as~\cite{c24,c25,c8}, utilize 3D convolutions to explicitly encode geometric information, which enhances accuracy but at the expense of high computational complexity. 

Iterative optimization-based methods, inspired by iterative refinement techniques in optical flow estimation such as RAFT~\cite{c27}, refine disparity estimates through multiple iterations. RAFT-Stereo~\cite{c19} is a pioneering work in this domain. RAFT employs a GRU (Gated Recurrent Unit) structure to iteratively optimize the disparity map, achieving a balance between high accuracy and efficiency. Building on this foundation, subsequent methods have further enhanced the iterative optimization process. The CREStereo~\cite{c28} adopts a hierarchical network design, updating the disparity map through a coarse-to-fine recursive approach to better recover complex image details. The IGEV-Stereo~\cite{c18} introduces adaptive correlation layers and geometry encoding volumes to improve the robustness and accuracy of the models. Similarly, DLNR~\cite{c10} employs an LSTM (Long-Short-Term Memory) structure to specifically optimize disparity estimates in high-frequency regions, thereby further bolstering the overall performance of the models.
\vspace{-0.2cm}
\section{Method}
\begin{figure}[htpb]
  \centering
  \vspace{-0.6cm}
  \includegraphics[width=\textwidth]{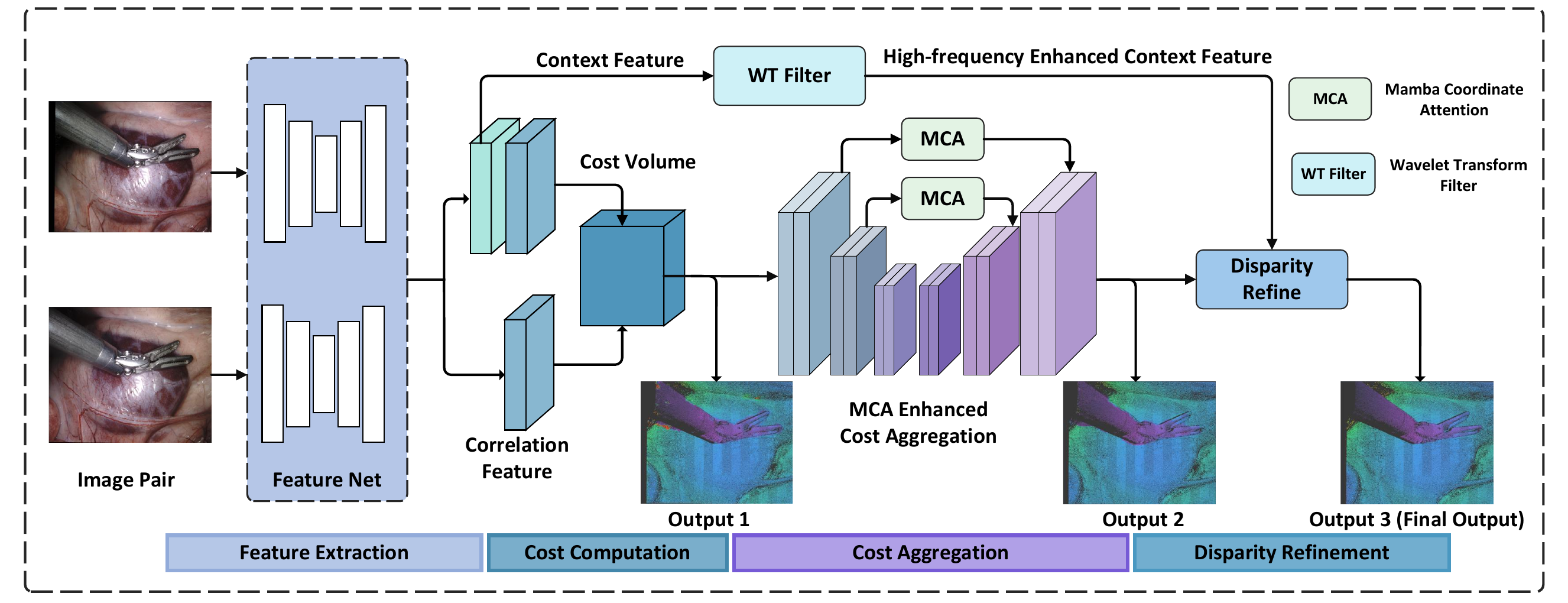}
\caption{
 Overall architecture of RRESM. The Feature Net adopts a U-Net-like structure, with a frozen encoder based on MobileNetV4 and a trainable decoder. A correlation cost volume is constructed using group-wise correlation. The MCA module is embedded within a simplified 3D CNN to enhance cost aggregation. The HFDO module takes deep features from the encoder as contextual information, applies a wavelet-based filter, and enhances high-frequency components in the disparity map. Details of the MCA and HFDO modules are shown in Fig.\ref{fig:module1} \ref{fig:module2}, respectively.
 }
   \vspace{-0.6cm}
  \label{fig:framework}
\end{figure}

\vspace{-0.2cm}
The overall architecture of RRESM is illustrated in Fig.~\ref{fig:framework}. The feature extraction backbone is based on MobileNetV4, providing a lightweight and efficient representation. The proposed MCA module is incorporated into a simplified 3D U-Net to perform cost aggregation in the feature space. Subsequently, the HFDO module is applied in the frequency domain to refine disparity predictions, particularly near boundary regions. Detailed descriptions of each component are provided in the following sections.

\subsection{Feature Extraction}

Multi-scale feature information is crucial for stereo matching: shallow features capture rich textures and fine geometric details, while deeper features capture higher-level semantics. To leverage both, we adopt a U-Net-like structure, where the downsampling branch is based on a frozen MobileNetV4, and the upsampling branch is trainable and gradually integrates multi-scale features to recover spatial resolution.

Formally, given a stereo pair \( I_l, I_r \in \mathbb{R}^{3 \times H \times W} \), we use the MobileNetV4 backbone~\cite{c22} as the downsampling path in our multi-scale feature extraction framework. In the upsampling path, features from resolutions of \( 1/16 \), \( 1/8 \), and \( 1/4 \) are progressively fused, producing final feature maps at \( 1/4 \) resolution: \( F_l, F_r \in \mathbb{R}^{C \times H/4 \times W/4} \). The feature \( F_l \) extracted from the downsampling branch is also used as context input for the subsequent disparity refinement module.

\subsection{Cost Volume Computation}
We employ a Group-wise Correlation cost volume~\cite{c8} to compute the matching cost between the extracted paired features. For the feature maps \( F_l \) and \( F_r \), we divide the channels into \( g_n \) groups, where \( g_n = 16 \) in this work. Within each group, the feature vectors from the left and right images are correlated via the inner product. The cost volume is then constructed by concatenating the correlation results across groups. The cost volume is defined as:

\begin{equation}
CV(g, d, x, y) = \frac{1}{C / g_n} \langle F_l^g(x, y), F_r^g(x - d, y) \rangle ,
\end{equation}

where \( CV \) denotes the cost volume, \( F_*^g \) represents the feature vectors within group \( g \) (\( g \in [0, g_n-1] \)), and \( \langle \cdot, \cdot \rangle \) denotes the inner product.

\subsection{3D Mamba Coordinate Attention Guided Cost Aggregation}
\begin{figure}[h]
  \centering
  \vspace{-0.4cm}
  \includegraphics[width=0.9\textwidth]{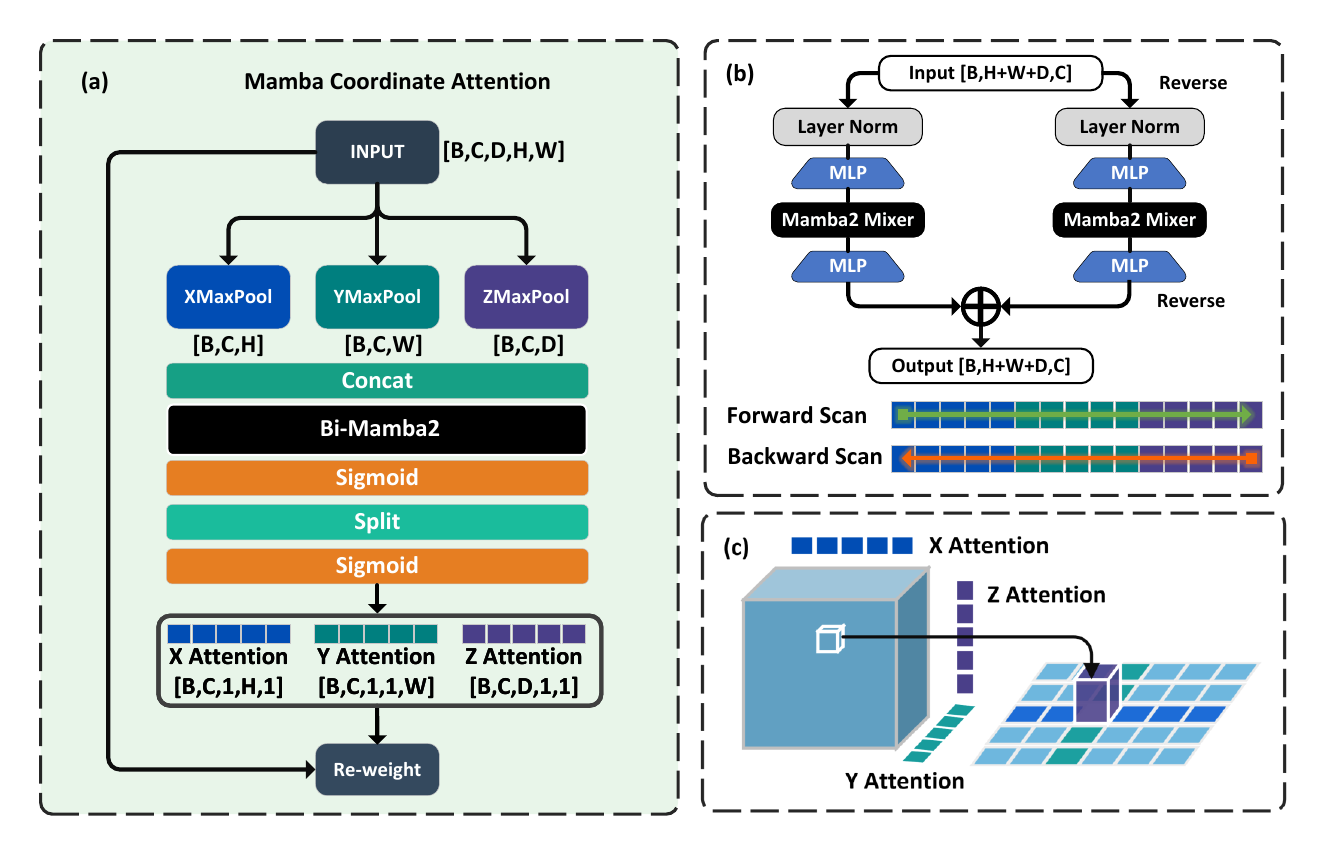}
  \vspace{-0.2cm}
\caption{
    \textbf{(a)} Architecture of the MCA module. Attention is computed independently along the \(H\), \(W\), and \(D\) dimensions, concatenated along the channel axis, and passed through a Bi-Mamba2 layer.  
    \textbf{(b)} Implementation of the Bi-Mamba2 layer in MCA. A bidirectional scan is performed over the concatenated axis-pooled features in both forward and backward directions.  
    \textbf{(c)} Re-weighting operation. The resulting position-sensitive attention maps assign a unique weight to each coordinate in the cost volume.
}
\vspace{-0.4cm}
  \label{fig:module1}
\end{figure}
\noindent
We propose the \textbf{3D Mamba Coordinate Attention (MCA)} module to perform efficient, position-sensitive cost aggregation, as illustrated in Fig.~\ref{fig:module1}. MCA enhances the conventional coordinate attention mechanism by embedding axis-specific positional information and explicitly modeling long-range dependencies across all three spatial dimensions—\textit{disparity (D)}, \textit{height (H)}, and \textit{width (W)}.

Recently, Mamba has emerged as a state-space model (SSM)-based alternative to transformers, offering strong sequence modeling capabilities with linear time and space complexity~\cite{dao2024transformers}. Unlike attention mechanisms that compute pairwise token interactions with quadratic cost, Mamba leverages structured state-space dynamics to capture global dependencies via fast kernel-based operations. This makes it well suited for real-time or high-resolution tasks, where transformers typically incur prohibitive computational overhead. Moreover, Mamba excels at capturing long-range spatial interactions, which aligns well with the requirements of 3D cost aggregation in stereo matching.

To incorporate spatial priors efficiently, we extend Mamba into a \textit{3D axis-aware attention mechanism}. Specifically, we treat the disparity dimension \( D \) as an additional spatial axis. Given a cost volume \( \mathcal{F} \in \mathbb{R}^{C \times D \times H \times W} \), we extract axis-aligned attention descriptors by applying global pooling over orthogonal spatial planes:

\begin{equation}
\begin{aligned}
\mathcal{Z}^x_c(x) &= \frac{1}{H \cdot D} \max_{y,z} \mathcal{F}(c, x, y, z), \\
\mathcal{Z}^y_c(y) &= \frac{1}{W \cdot D} \max_{x,z} \mathcal{F}(c, x, y, z), \\
\mathcal{Z}^z_c(z) &= \frac{1}{W \cdot H} \max_{x,y} \mathcal{F}(c, x, y, z),
\end{aligned}
\end{equation}
where \(\mathcal{Z}^x\), \(\mathcal{Z}^y\), and \(\mathcal{Z}^z\) denote the axis-specific pooled descriptors. These vectors are passed through a sigmoid activation and concatenated as:

\begin{equation}
A_{\text{cat}} = \text{concat}\left[ \delta(\mathcal{Z}^x), \delta(\mathcal{Z}^y), \delta(\mathcal{Z}^z) \right],
\end{equation}

\noindent
where \(\delta(\cdot)\) is the sigmoid function.

The original Mamba (Mamba1)~\cite{gu2023mamba} performs unidirectional sequence modeling along a given axis, which may lead to directional bias. To enhance symmetry and spatial coherence, we employ the Mamba2, which applies bidirectional sequential modeling along each axis. This enables the model to aggregate contextual cues both forward and backward across disparity, height, and width dimensions, which is particularly beneficial near fuzzy anatomical boundaries.

The refined attention maps are obtained as:

\begin{equation}
A_{\text{refined}} = \text{Mamba2}(A_{\text{cat}}), \quad A_x, A_y, A_z = \text{split}(A_{\text{refined}}),
\end{equation}

\noindent
and applied to the original cost volume by broadcasted element-wise multiplication:

\begin{equation}
\mathcal{F}_{\text{refined}} = \mathcal{F} \odot A_x \odot A_y \odot A_z,
\end{equation}
\noindent
where \(\odot\) denotes Hadamard (element-wise) product. The refined feature volume \( \mathcal{F}_{\text{refined}} \) is then passed to a lightweight 3D U-Net cost aggregation network, improving global spatial reasoning while maintaining low computational complexity.

\subsection{High Frequency Disparity Optimization}
\begin{figure}[htbp]
  \centering
  \vspace{-0.4cm}
  \includegraphics[width=0.9\textwidth]{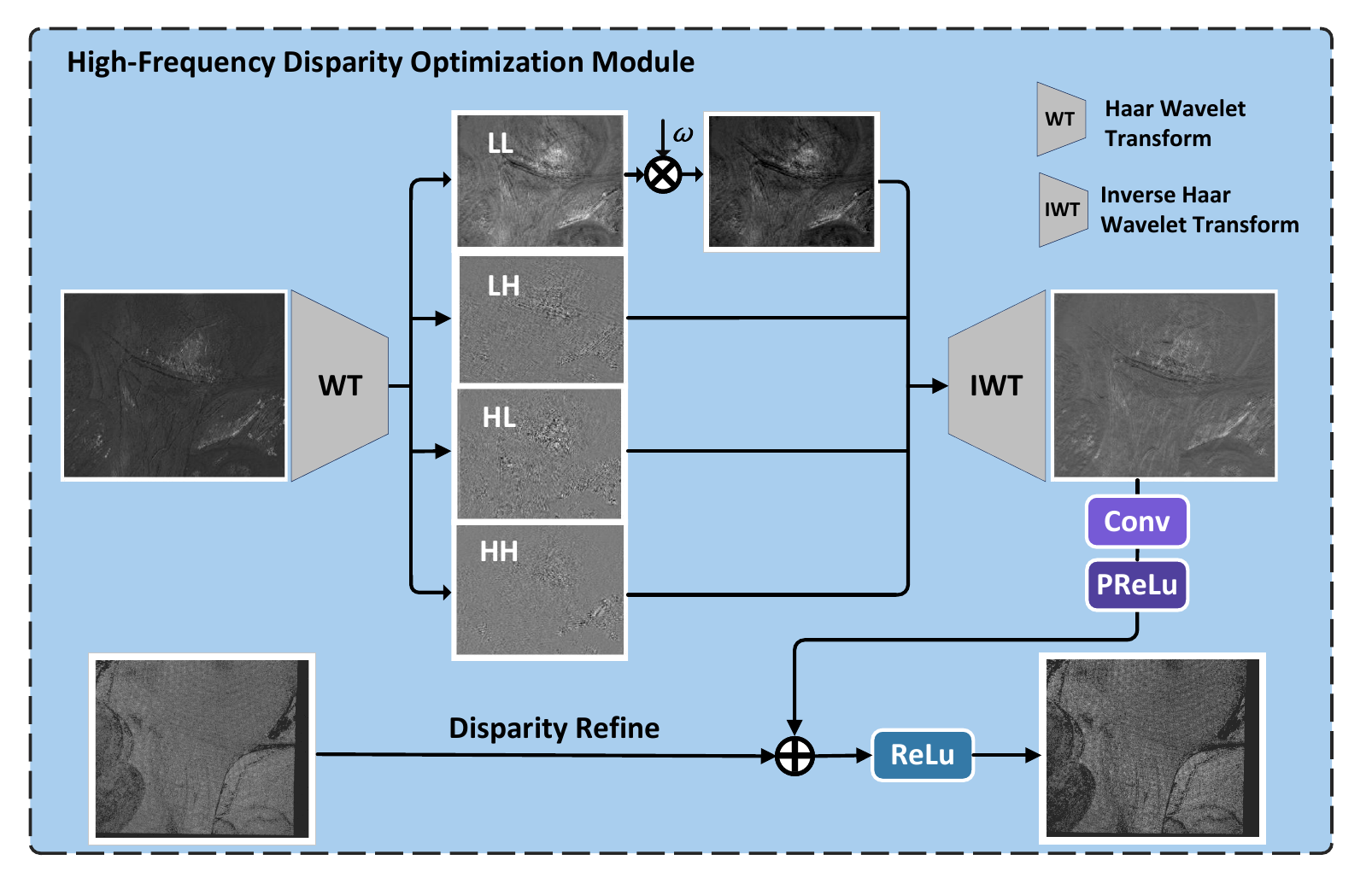}
  \vspace{-0.2cm}
\caption{
     Wavelet Transform Refine module. The context feature map is decomposed into low-frequency (LL) and high-frequency (LH, HL, HH) components. The LL components are attenuated by a parameter \(w\).
}
\vspace{-0.4cm}
  \label{fig:module2}
\end{figure}
To further improve disparity estimation near fuzzy boundaries, we introduce a wavelet-based module, \textbf{HFDO}, which refines the disparity map in the frequency domain. This module decomposes the context feature \( F_s \), extracted from \( F_l \), into low-frequency components (LL) and high-frequency components in the horizontal (LH), vertical (HL), and diagonal (HH) directions via a Haar wavelet transform. To emphasize high-frequency structures, the low-frequency component is attenuated by a scaling factor \(\omega\), and the inverse wavelet transform is subsequently applied to reconstruct a refined context feature. This refined representation specifically enhances high-frequency details, improving disparity prediction in regions with ambiguous or fuzzy tissue boundaries.

The context features are derived as follows:
\begin{equation}
F_s = \text{ReLU}(\text{Linear}(F_l)),
\end{equation}
where \(\text{Conv}\) denotes a 2D \(3\times 3\) convolution.  
The 2D Haar wavelet convolution kernels used are~\cite{c30}:

\begin{equation}
h_{LL} = \frac{1}{2}
\begin{bmatrix}
1 & 1 \\
1 & 1
\end{bmatrix},\ 
h_{LH} = \frac{1}{2}
\begin{bmatrix}
1 & -1 \\
1 & -1
\end{bmatrix},\ 
h_{HL} = \frac{1}{2}
\begin{bmatrix}
1 & 1 \\
-1 & -1
\end{bmatrix},\ 
h_{HH} = \frac{1}{2}
\begin{bmatrix}
1 & -1 \\
-1 & 1
\end{bmatrix}.
\end{equation}

The decomposition of the feature map $F_s$ is performed as follows:

\begin{equation}
F_s^{ll} = \omega \cdot (h_{LL} * F_s),\ F_s^{lh} = h_{LH} * F_s,\ 
F_s^{hl} = h_{HL} * F_s,\ F_s^{hh} = h_{HH} * F_s,
\end{equation}
where \( \omega \) is the attenuation factor for the low-frequency components. The feature map is then reconstructed using the inverse wavelet transform:
\begin{equation}
F_s^{\text{filtered}} = \text{IWT}(F_s^{ll}, F_s^{lh}, F_s^{hl}, F_s^{hh}).
\end{equation}

The filtered semantic features \( F_s^{\text{filtered}} \) are projected into the disparity space through a \(\text{Conv}\) layer with a PReLU~\cite{he2015delving} activation function to generate the optimized disparity map. Finally, the optimized disparity map is processed using a ReLU activation function to ensure that the disparity values are non-negative:
\begin{equation}
    D_{refined} = \text{ReLU}(D + \text{PReLU}(\text{Conv}(F_s^{filtered}))),
\end{equation}

\subsection{Loss Function}
We utilize the \(smooth\ L_1\) loss(Eq. \ref{sml1}) to quantify the discrepancy between the predicted disparity map and the ground-truth disparity map. Specifically, we compute the loss at three critical stages of the disparity estimation pipeline: prior to cost aggregation (\(d_{f}\)), before disparity optimization (\(d_{cg}\)), and at the final output stage (\(d_{dr}\)).:
\begin{equation}
    L_{smooth}(x) =  \begin{cases}
0.5 x^2 & \text{if } |x| < 1, \\
|x| - 0.5 & \text{otherwise}.
\label{sml1}
\end{cases}
\end{equation}
The total loss is formulated as:
\begin{equation}
\begin{split}
    \text{Loss} = w_1 \cdot \|d_f - d_{gt}\|_1 + w_2 \cdot \|d_{cg} - d_{gt}\|_1 +
w_3 \cdot \|d_{dr} - d_{gt}\|_1,
\end{split}
\end{equation}
where \(w_1\), \(w_2\), and \(w_3\) are the weights assigned to the losses at each respective stage(we set \(w_1=w_2=w_3=1/3\)).

\section{Experiments}
\subsection{Datasets}
We evaluate our method on two publicly available dataset:

\vspace{0.2cm}
\noindent
The \textbf{SCARED dataset}~\cite{c20} is a public laparoscopy dataset from the MICCAI 2019 Endovis Challenge, captured using the da Vinci Xi surgical robot. It consists of porcine peritoneal images with a resolution of \(1024 \times 1280\). The dataset includes 7 training subsets and 2 test subsets. Due to significant calibration errors in datasets 4 and 5, we discarded these subsets and trained our model using the remaining training data (14,714 image pairs). The dataset 8 and 9 are testing data. We adopt the official evaluation metric: Mean Absolute Error (MAE) in mm of the depth map.

\vspace{0.2cm}
\noindent
The \textbf{SERV-CT dataset}~\cite{c29} contains 16 pairs of porcine peritoneal stereo images with a resolution of \(720 \times 576\). We use all 16 pairs for testing. We evaluate performance using the Mean Absolute Error (MAE) in pixels, the percentage of pixels with a disparity error greater than n pixels (Bad-n), and the percentage of pixels with an error greater than 3 pixels and greater than 5\% of the ground-truth value (D1).

\subsection{Implementation Details}
Our implementation is based on the PyTorch framework. We utilized the Adam optimizer with $\beta_1 = 0.9$ and $\beta_2 = 0.999$. The learning rate is set to $1 \times 10^{-4}$. For data augmentation, we applied random cropping to resize the training images to $256 \times 512$. The model was trained for 100 epochs on the SCARED dataset. For evaluation, we tested the model on datasets 8 and 9 of the SCARED dataset, as well as the entire SERV-CT dataset. All experiments were conducted on an Ubuntu 22.04 system with 4 Nvidia RTX 2080Ti GPUs.

\subsection{Evaluations on SCARED and SERV-CT}
We evaluated our algorithm against a set of state-of-the-art stereo matching methods, including both those designed for natural images and those specifically developed for endoscopic data. In particular, GwcNet~\cite{c8}, PSMNet~\cite{c25}, RAFT~\cite{c9}, IGEV~\cite{c18}, DLNR~\cite{c10}, and Selective-Stereo~\cite{c31} are representative natural image methods. Among them, DLNR and Selective-Stereo explicitly focus on optimizing high-frequency details in disparity estimation. Except for PSMNet, which we report using results cited from~\cite{c34,c35}, all other models were retrained on the SCARED dataset under a consistent training pipeline.
\begin{table}[htpb]
\centering
\vspace{-0.8cm}
\caption{Results of comparison experiments on the SERV-CT dataset.}
\label{table2}
\resizebox{0.82\columnwidth}{!}{
\begin{tabular}{lccccc}
\toprule
\textbf{Model} & \textbf{MAE(pixel)$\downarrow$} & \textbf{D1$\downarrow$} & \textbf{Bad1$\downarrow$} & \textbf{Bad2$\downarrow$} & \textbf{Bad3$\downarrow$}\\
\midrule
GwcNet  & 5.813& 33.92\%& 79.37\%& 56.32\%& 40.93\% \\
PSMNet*& 6.355 & - & \underline{70.88\%} & \underline{49.80\%} &\underline{33.87\%} \\
RAFT  & 21.500 & 87.50\% & 95.80\% & 91.80\% & 88.50\% \\
IGEV  & 32.895& 92.91\%& 98.28\%& 96.53\%& 94.76\% \\
DLNR  & 22.030& 88.97\%& 96.21\%& 92.61\%& 89.13\% \\
Selective-Stereo& 24.103& 86.36\%& 95.68\%& 91.33\%& 87.01\%\\
MSDESIS-full & \underline{5.742} & \underline{28.26\%} & 78.81\%& 55.44\%& 37.79\%\\
MSDESIS-light & 7.065& 41.35\%& 81.09\%& 63.17\%&49.22\%\\
\textbf{RRESM(ours)} & \textbf{2.367} & \textbf{14.34\%} & \textbf{58.59\%} & \textbf{31.52\% }& \textbf{18.49\%}\\
\bottomrule
\end{tabular}
 }
\vspace{-0.4cm}
\end{table}

\begin{table}[htpb]
\centering
\vspace{-0.6cm}
\caption{MAE (unit: mm) for the SCARED Test Set. Each test set comprises 5 keyframes (kf1--kf5). Lower values indicate better performance. Methods marked with (*) reference results from other papers. \textbf{Bold} indicates the best performance, and \underline{underlined} values indicate the second-best.}
\label{table1}
\resizebox{0.99\columnwidth}{!}{
\begin{tabular}{lccccccccccc}
\toprule
\textbf{Method} & \multicolumn{5}{c}{\textbf{Dataset 8}} & \multicolumn{5}{c}{\textbf{Dataset 9}} & \textbf{ } \\
\cmidrule(lr){2-6} \cmidrule(lr){7-11}
& \textbf{kf1} & \textbf{kf2} & \textbf{kf3} & \textbf{kf4} & \textbf{kf5} & \textbf{kf1} & \textbf{kf2} & \textbf{kf3} & \textbf{kf4} & \textbf{kf5} & \textbf{Avg.} \\
\midrule
GwcNet & 9.07 & 2.92 & \underline{1.43} & \textbf{1.70} & 1.20 & \underline{3.86} & 1.13 & 2.93 & 1.99 & 0.61 & 2.68 \\
PSMNet*& 8.96 & 2.77 & \textbf{1.43} & 1.83 & 1.42 & 3.99 & \underline{1.08} & 2.82 & 1.95 & \underline{0.56} & 2.68\\
RAFT & 7.92 & 2.36 & 1.72 & 2.16 & 1.88 & 4.36 & \underline{1.08} & 2.87 & 1.98 & 1.54 & 2.79 \\
IGEV & 7.95 & 2.39 & 1.69 & 2.26 & 1.93 & 3.89 & 1.17 & 2.89 & 2.50 & 1.25 & 2.79 \\
DLNR & 7.84 & 2.52 & 2.65 & 3.20 & 3.41 & 5.37 & 1.91 & 4.32 & 4.82 & 10.25 & 4.63\\
Selective-Stereo& \textbf{7.53} & \underline{2.40} & 1.66 & 2.31 & 2.26 & \textbf{3.86} & \textbf{0.95} & 2.83 & \underline{1.77} & 1.01 & 2.66\\
MSDESIS-full& 10.14 & 3.46 & 3.30 & 3.88 & 2.09 & 5.42 & 1.63 & 3.37 & \textbf{1.76} & 0.63 & 3.57 \\
MSDESIS-light & 11.45 & 15.91 & 7.72 & 10.67 & 9.84 & 98.39 & 130.80 & 66.57 & 101.80 & 203.40 & 65.67 \\
DCStereo*& 8.38 & 2.77 & 1.54 & \underline{1.82}& \underline{1.19}& 3.99 & 1.09 & \textbf{2.64} & 2.09 & 1.04 & \underline{2.65} \\
\textbf{RRESM(ours)} & \underline{7.62} & \textbf{2.31} & 1.66 & 2.28 & \textbf{1.02} & 4.42 & 1.16 & \underline{2.80} & 2.12 & \textbf{0.47} & \textbf{2.59} \\
\bottomrule
\end{tabular}
}
\vspace{-0.6cm}
\end{table}

In contrast, MSDESIS~\cite{c21} and DCStereo~\cite{c34} are methods tailored for endoscopic stereo matching. For DCStereo, we directly cited its results on the SCARED dataset, as their evaluation protocol is consistent with ours. However, since DCStereo was not evaluated on the SERV-CT dataset and its implementation is not publicly available, we omitted it from SERV-CT comparisons. As for MSDESIS, we used the pre-trained model provided by the authors. Note that our method uses a maximum disparity of 192, while MSDESIS adopts 320, which may account for slight differences between our reproduced results and those reported in their original paper. 

\begin{figure}[h]
  \centering
  \vspace{-0.8cm}
  \includegraphics[width=\textwidth]{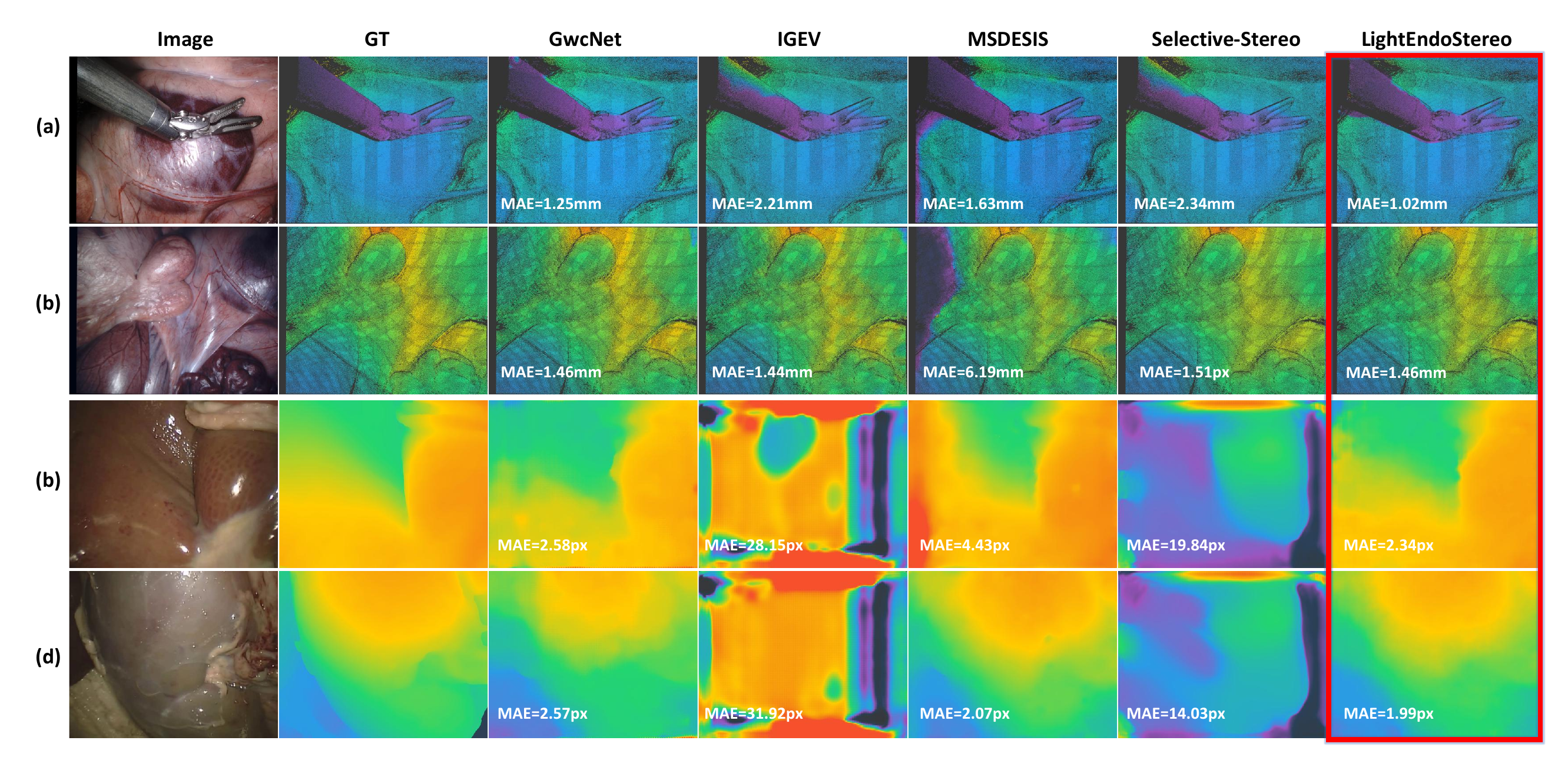}
  \vspace{-0.8cm}
\caption{Visualization of disparity estimation on the SCARED and SERV-CT datasets. (a) and (b) are from SCARED, while (c) and (d) are from SERV-CT. In cases with limited depth variation, such as (b), most methods perform similarly. However, in high-frequency regions like surgical tool-tissue boundaries in (a), RRESM yields more accurate depth predictions. On SERV-CT, our method also delivers competitive results. \textbf{Note:} MAE is measured in millimeters (mm) for SCARED (with ground-truth depth) and in pixels for SERV-CT (with ground-truth disparity).}
  \vspace{-0.6cm}
  \label{result1}
\end{figure}

The quantitative results on the SCARED and SERV-CT datasets are presented in Table~\ref{table1} and Table~\ref{table2}, respectively. Sample visualizations are shown in Fig.~\ref{result1}. On the SCARED dataset, RRESM achieves the best average MAE of 2.59\,mm, outperforming both natural image and endoscopy-specific baselines. On the SERV-CT dataset, our method also achieves superior performance across all metrics, including the lowest MAE (2.367 pixels), lowest Bad-n error rates, and a significantly lower D1 score. These results highlight RRESM's robustness in handling challenging tissue boundaries and its strong generalization ability to unseen clinical data.

\subsection{Runtime Evaluation}
\begin{table}[htpb]
\centering
\vspace{-0.8cm}
\caption{Quantitative results of model computational performance.}
\label{table3}
\begin{tabular}{lccc}
\toprule
\textbf{Model} & \textbf{Params(M)} & \textbf{FLOPs(T)} & \textbf{Runtimes(ms)} \\
\midrule
GwcNet & 6.43 & \underline{2.466} & \underline{26.4}\\
PSMNet*& \underline{3.672} & - & 225\\
RAFT & 11.11 & 3.757 & 468.6\\
IGEV& 12.5 & 3.288 & 434.5\\
Selective-Stereo& 13.141& 4.062& 516.52\\
DCStereo*& \textbf{3.404}& -& 191\\
\textbf{RRESM(ours)} & 11.094 & \textbf{0.846} & \textbf{23.38}\\
\bottomrule
\end{tabular}
\vspace{-0.4cm}
\end{table}
To evaluate the runtime performance of our model, we conducted tests using a single Nvidia RTX 2080Ti GPU. We sampled 100 images ($1280 \times 1024$) from dataset 8 of the SCARED for testing and used the average inference time as the evaluation metric. The results are shown in Table \ref{table3}. Although the MSDESIS-Light model has the smallest footprint, its matching accuracy is poor. In contrast, RRESM achieved a real-time matching speed of 23.38 ms per frame.

\subsection{Ablation Study}

\begin{table}[h]
\centering
\vspace{-0.6cm}
\caption{Ablation Study Results. The average MAE on the SCARED dataset is used as the accuracy evaluation metric.}
\label{table4}
\begin{tabular}{lccccl}
\toprule
\textbf{MCA}& \textbf{HFDO}& \textbf{MAE(mm)}& \textbf{Runtime(ms)}& \textbf{Params(M)} &\textbf{FLOPs(T)}\\
\midrule
& & 2.718& 25.412ms& 13.047&1.755 \\
 & $\checkmark$& 2.645& 25.972ms& 13.077&1.757\\
 \checkmark & & 2.621& 22.655ms& 11.064&0.843\\
 \checkmark& $\checkmark$& 2.592& 23.386ms& 11.094&0.846\\
 \bottomrule
\end{tabular}
\vspace{-0.2cm}
\end{table}
We conducted ablation studies to evaluate the contributions of the proposed \textbf{3D Mamba Coordinate Attention (MCA)} and \textbf{High-Frequency Disparity Optimization (HFDO)} modules. MCA is designed to improve cost aggregation efficiency by modeling axis-specific long-range dependencies, while HFDO focuses on refining depth estimates in regions with high-frequency disparity transitions, such as tissue-tool boundaries.
\begin{figure}[h]
  \centering
  \vspace{-0.4cm}
  \includegraphics[width=0.8\textwidth]{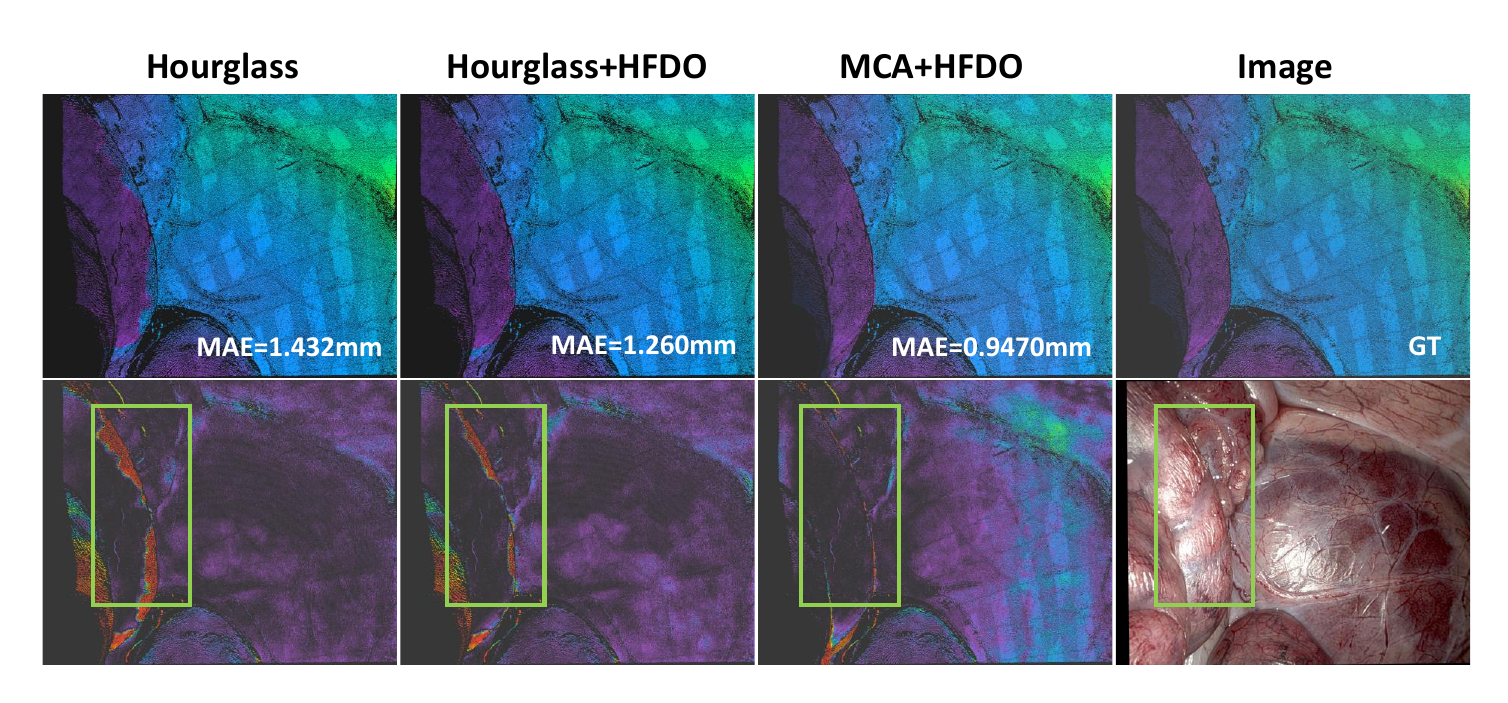}
  \vspace{-0.4cm}
    \caption{Visualization of disparity optimization in high-frequency regions. The top row shows the depth maps, while the bottom row displays the corresponding error maps.}
    \vspace{-0.4cm}
  \label{result2}
\end{figure}

The experiments were performed on the SCARED dataset using our 3D U-Net-based architecture. To assess MCA, we replaced it with a conventional stacked 3D Hourglass network~\cite{c8}, which is commonly used in existing stereo matching pipelines. For evaluating HFDO, we simply removed the wavelet-based refinement module from the full pipeline. We report the Mean Absolute Error (MAE) in millimeters, runtime per frame, model parameters, and FLOPs, as summarized in Table~\ref{table4}.

Results show that both modules individually contribute to improved accuracy. Incorporating MCA alone reduces MAE from 2.718 to 2.621\,mm while significantly lowering model complexity. Adding HFDO further improves performance to 2.592\,mm. Notably, the complete model achieves the best trade-off between accuracy and efficiency. As visualized in Fig.~\ref{result2}, HFDO particularly enhances boundary precision by reducing local matching errors in high-frequency regions. These results validate the effectiveness and complementary nature of MCA and HFDO within the overall architecture.

\section{Discussion and Conclusions}
We present RRESM, a real-time stereo matching framework tailored for endoscopic imagery. By integrating a 3D Coordinate Attention mechanism with the Mamba block, RRESM enables efficient and lightweight cost aggregation while capturing long-range dependencies. The High-Frequency Disparity Optimization module further enhances disparity accuracy near anatomical edges through wavelet-based refinement. Extensive experiments on the SCARED and SERV-CT datasets show that RRESM achieves state-of-the-art performance and generalizes well to unseen domains, running at 42 FPS on high-resolution inputs.

Despite these promising results, our method has several limitations. Its performance may degrade under extreme illumination variation, specular reflections, or endoscope lens occlusion. Additionally, the current pipeline is limited to stereo input and assumes spatial rectification. In future work, we will (1) explore uncertainty-aware disparity refinement, (2) extend RRESM to multi-view and monocular depth estimation, and (3) validate its robustness across diverse anatomical sites and surgical conditions.

\bibliographystyle{splncs04}
\bibliography{ref}
\end{document}